\documentclass[twocolumn,showpacs,preprintnumbers,amsmath,amssymb,prl,superscriptaddress,nofootinbib]{revtex4}

\usepackage{graphicx}
\usepackage{dcolumn}
\usepackage[loose]{subfigure}      
\usepackage{bm}
\newcommand*{\INFNPi}{\affiliation{INFN Sezione di Pisa$^{a}$; Dipartimento di Fisica$^{b}$ dell'Universit\`a, Largo B.~Pontecorvo~3, 56127 Pisa, Italy}}
\newcommand*{\INFNGe}{\affiliation{INFN Sezione di Genova$^{a}$; Dipartimento di Fisica$^{b}$ dell'Universit\`a, Via Dodecaneso 33, 16146 Genova, Italy}}
\newcommand*{\INFNPv}{\affiliation{INFN Sezione di Pavia$^{a}$; Dipartimento di Fisica Nucleare e Teorica$^{b}$ dell'Universit\`a, Via Bassi 6, 27100 Pavia, Italy}}
\newcommand*{\INFNRm}{\affiliation{INFN Sezione di Roma$^{a}$; Dipartimento di Fisica$^{b}$ dell'Universit\`a ``Sapienza'', Piazzale A.~Moro, 00185 Roma, Italy}}
\newcommand*{\INFNLe}{\affiliation{INFN Sezione di Lecce$^{a}$; Dipartimento di Fisica$^{b}$ dell'Universit\`a, Via per Arnesano, 73100 Lecce, Italy}}
\newcommand*{\ICEPP} {\affiliation{ICEPP, University of Tokyo 7-3-1 Hongo, Bunkyo-ku, Tokyo 113-0033, Japan }}
\newcommand*{\UCI}   {\affiliation{University of California, Irvine, CA 92697, USA}}
\newcommand*{\KEK}   {\affiliation{KEK, High Energy Accelerator Research Organization 1-1 Oho, Tsukuba, Ibaraki 305-0801, Japan}}
\newcommand*{\PSI}   {\affiliation{Paul Scherrer Institut PSI, CH-5232 Villigen, Switzerland}}
\newcommand*{\Waseda}{\affiliation{Research Institute for Science and Engineering, Waseda~University, 3-4-1 Okubo, Shinjuku-ku, Tokyo 169-8555, Japan}}
\newcommand*{\BINP}   {\affiliation{Budker Institute of Nuclear Physics, 630090 Novosibirsk, Russia}}
\newcommand*{\JINR}   {\affiliation{Joint Institute for Nuclear Research, 141980, Dubna, Russia}}
\newcommand*{\ETHZ}   {\affiliation{Swiss Federal Institute of Technology ETH, CH-8093 Z\" urich, Switzerland}}
\newcommand*{\BR}     { {\cal B} }
\newcommand*{\meg}            {\mu \to e \gamma}
\newcommand*{\megsign}        {\mu^+ \to e^+ \gamma}

\newcommand*{\michelsign}        {\mu^+ \to e^+ \nu\bar{\nu}}
\newcommand*{\egamma}         {E_{\gamma}}
\newcommand*{\epositron}      {E_e}
\newcommand*{\tegamma}        {t_{e\gamma}}
\newcommand*{\Thetaegamma}    {\Theta_{e\gamma}}

\newcommand*{\thetaegamma}    {\theta_{e\gamma}}
\newcommand*{\phiegamma}      {\phi_{e\gamma}}

\newcommand*{\nsig}           {N_{\rm sig}}
\newcommand*{\nrd}            {N_{\rm RMD}}
\newcommand*{\nbg}            {N_{\rm BG}}
\newcommand*{\nobs}           {N_{\rm obs}}

\begin{document}

\title{New limit on the lepton-flavour violating decay $\mu^{+} \to e^{+} \gamma$}

\author{J.~Adam}               \PSI\ETHZ
\author{X.~Bai}                \ICEPP
\author{A.~M.~Baldini$^{a}$}   \INFNPi
\author{E.~Baracchini}         \UCI
\author{C.~Bemporad$^{ab}$}    \INFNPi
\author{G.~Boca$^{ab}$}        \INFNPv
\author{P.~W.~Cattaneo$^{a}$}  \INFNPv
\author{G.~Cavoto$^{a}$}       \INFNRm
\author{F.~Cei$^{ab}$}         \INFNPi
\author{C.~Cerri$^{a}$}        \INFNPi
\author{A.~de~Bari$^{ab}$}      \INFNPv
\author{M.~De~Gerone$^{ab}$}   \INFNGe
\author{T.~Doke}               \Waseda
\author{S.~Dussoni$^{ab}$}     \INFNGe
\author{J.~Egger}              \PSI
\author{K.~Fratini$^{ab}$}     \INFNGe
\author{Y.~Fujii}              \ICEPP
\author{L.~Galli$^{ab}$}        \INFNPi
\author{G.~Gallucci$^{ab}$}     \INFNPi
\author{F.~Gatti$^{ab}$}        \INFNGe
\author{B.~Golden}              \UCI
\author{M.~Grassi$^{a}$}        \INFNPi
\author{D.~N.~Grigoriev}        \BINP
\author{T.~Haruyama}            \KEK
\author{M.~Hildebrandt}         \PSI
\author{Y.~Hisamatsu}           \ICEPP
\author{F.~Ignatov}             \BINP
\author{T.~Iwamoto}             \ICEPP
\author{P.-R.~Kettle}           \PSI
\author{B.~I.~Khazin}           \BINP
\author{O.~Kiselev}             \PSI
\author{A.~Korenchenko}         \JINR
\author{N.~Kravchuk}            \JINR
\author{A.~Maki}                \KEK
\author{S.~Mihara}              \KEK
\author{W.~Molzon}              \UCI
\author{T.~Mori}                \ICEPP
\author{D.~Mzavia\stepcounter{footnote}}              \thanks{deceased} \JINR
\author{H.~Natori}              \ICEPP\PSI
\author{D.~Nicol\`o$^{ab}$}     \INFNPi
\author{H.~Nishiguchi}          \KEK
\author{Y.~Nishimura}           \altaffiliation[Present address:]{\ Kamioka Observatory, ICRR, University of Tokyo 456 Higashi-Mozumi, Kamioka-cho, Hida-city, Gifu 506-1205, Japan}       \ICEPP
\author{W.~Ootani}              \ICEPP
\author{M.~Panareo$^{ab}$}      \INFNLe
\author{A.~Papa}                \PSI
\author{R.~Pazzi$^{ab}$}        \thanks{deceased}\INFNPi
\author{G.~Piredda$^{a}$}       \INFNRm
\author{A.~Popov}               \BINP
\author{F.~Renga$^{a}$}         \INFNRm\PSI
\author{S.~Ritt}                \PSI
\author{M.~Rossella$^{a}$}      \INFNPv
\author{R.~Sawada}              \ICEPP
\author{F.~Sergiampietri$^{a}$}\INFNPi
\author{G.~Signorelli$^{a}$} \INFNPi
\author{S.~Suzuki}              \Waseda
\author{F.~Tenchini$^{ab}$}     \INFNPi
\author{C.~Topchyan}            \UCI
\author{Y.~Uchiyama}            \ICEPP\PSI
\author{R.~Valle$^{ab}$}        \altaffiliation[Present address:]{\ Lames Holding S.r.l., 16043 Chiavari, Italy}    \INFNGe
\author{C.~Voena$^{a}$}         \INFNRm
\author{F.~Xiao}                \UCI
\author{S.~Yamada}              \KEK
\author{A.~Yamamoto}            \KEK
\author{S.~Yamashita}           \ICEPP
\author{Yu.~V.~Yudin}           \BINP
\author{D.~Zanello$^{a}$}       \INFNRm

\collaboration{MEG Collaboration}
\noaffiliation

\date{\today}

\begin{abstract}
We present a new result based on an analysis of the data collected by the MEG detector at
the Paul Scherrer Institut in 2009 and 2010, in search of the lepton flavour
violating decay $\mu^+ \to e^+ \gamma$. 
The likelihood analysis of the combined data sample, which
corresponds  to a total of $1.8 \times 10^{14}$ muon decays, 
gives a 90\% C.L. upper limit of $2.4 \times 10^{-12}$ on the branching ratio 
of the $\mu^+ \to e^+ \gamma$ decay, 
constituting the most stringent limit on the existence of this decay to date.

\end{abstract}


\maketitle
The lepton flavour violating (LFV) decay $\meg$ is forbidden within
the standard model of elementary particles (SM). Even with the introduction 
of neutrino masses and mixing SM predicts an immeasurably small
branching ratio ($\BR \lesssim$ 10$^{-51}$) for this decay. Conversely new physics 
scenarios beyond
SM, such as  supersymmetric grand unified theories  or theories with extra dimensions,
predict branching ratios in the $10^{-12}$ to 10$^{-14}$ range~\cite{barbieri, hisano,
LFV-EPC}. This is close to the present limit set by the MEGA
experiment~\cite{MEGA}, $\BR \leq 1.2 \times 10^{-11}$, which places one
of the most stringent constraints on the formulation of such theories. 
Observation of $\meg$ therefore would be an unambiguous signature of new
physics, while improvements on the existing limit would stringently constrain 
many of the new physics scenarios beyond SM.

The MEG experiment~\cite{proposal, MEG-NPB} covers a 10\% solid
angle, centred around a thin muon stopping target ($205\,\mu$m-thick polyethylene) and is composed of a
positron spectrometer and a photon detector
in search of back-to-back, monoenergetic,
time coincident photons and positrons from the two-body $\megsign$ decays.
The positron spectrometer consists of
a set of drift chambers (DC)~\cite{DC} and scintillation timing counters (TC)~\cite{TC} 
located inside a superconducting solenoid with a gradient field ~\cite{COBRA} 
along the beam axis, ranging from 1.27 Tesla at the centre to 0.49 Tesla at either end.
The photon detector~\cite{LXe}, located outside of the solenoid, is a
homogeneous volume (900~$\ell$) of liquid xenon (LXe) viewed by 846
UV-sensitive photo-multiplier tubes (PMTs) submerged in the liquid.
The spectrometer measures the positron momentum vector and timing, while
the LXe detector is used to reconstruct the $\gamma-$ray
energy as well as the position and time of its first interaction in LXe.
All the signals are individually digitized by in-house designed
waveform digitizers based on the multi-GHz domino
ring sampler chip (DRS)~\cite{DRS}.
The  PSI $\pi {\rm E} 5$  beam line is used to stop
$3 \times 10^{7 }$ positive muons per second  in the target.
 The residual polarization of the decaying muons along the beam axis was measured to be 
 $\langle P \rangle = -0.89 \pm 0.04$.
The background to $\megsign$ decay comes either from radiative muon
decays  $\michelsign \gamma$ (RMD) in which the neutrinos carry away little
energy or from an accidental coincidence of an energetic positron
from a normal Michel decay with a $\gamma-$ray coming from
RMD, Bremsstrahlung or positron annihilation-in-flight. 
The accidental coincidences are the dominant background in this experiment.

The MEG detector response, resolutions and stability
are constantly monitored and calibrated. 
The photomultipliers (PMTs) of the LXe detector are calibrated daily by 
LEDs and $\alpha$-sources immersed in the liquid~\cite{alpha}.
The energy scale
and resolutions of the LXe detector are measured over the energy range of $4.43$ to
$129.4$~MeV using $\gamma-$rays from a radioactive Am/Be source,
$(p,\gamma)$-reaction using a dedicated Cockcroft-Walton accelerator
(CW)~\cite{CW}, and $\pi^-p$ charge exchange and radiative capture reactions
(CEX).
A 9\,MeV-$\gamma$ line from the capture in nickel of neutrons from 
a pulsed and triggerable deuteron-deuteron neutron generator 
allows one to check the stability of the LXe detector even during data-taking. 
The relative time between the TC and LXe detector is monitored
using RMD and 2$\gamma$-events from 
$^{11}_{\,\,\,5}\mathrm{B}(p,2\gamma)^{12}_{\,\,\,6}\mathrm{C}$ 
reactions.

The $\megsign$  trigger requires the
presence of a high energy $\gamma-$ray in the LXe detector and a hit
on the timing counters within a 20~ns window together with an approximate
back-to-back topology. Pre-scaled monitoring and calibration
triggers are also recorded.
A more detailed description of the MEG detector
can be found in Ref.~\cite{MEG-NPB}.

The results presented in this paper are based on data collected in 2009 and 2010
(for a total of  $1.8 \times 10^{14}~\mu^+$-decays
in the target); the 2010 statistics are about twice that of 2009.
All sub-detectors were running stably during these periods.
The 2008 data~\cite{MEG-NPB} are not used in this analysis 
because of their limited statistics and detector performance. 
In 2010 a DRS upgrade resulted in an improvement in the time resolution while 
an increase in noise in the DC, due to a deterioration of the HV power supplies, caused 
slightly worse positron tracking resolutions. 

We adopted a likelihood analysis method combined with a blind procedure on examining the data: events close to the signal region were kept hidden (blind region) 
until all the analysis procedures had been completely defined. 
The probability density functions (PDFs) needed for the likelihood analysis were
constructed using the events outside of the blind region (side-bands).

Several improvements to the analysis have been introduced since 
the presentation of the preliminary result based on the 2009 data~\cite{ICHEP}
and also implemented in the current 2009, as well as in our 2010 analyses.
These improvements include a new alignment technique for the DC system; 
an improved experimental evaluation of the spectrometer performances; 
a better understanding of the gradient magnetic field; 
improvement in the relative alignment of the photon detector and the positron spectrometer by means of cosmic ray muons; 
adoption of a more commonly used statistical method (profile likelihood); 
a constraint on the background rates in the likelihood analysis from the data 
in the side-bands.

The kinematic variables used to identify the $\megsign$ decays are the 
$\gamma$-ray and $e^+$ energies ($\egamma$, $\epositron$), their relative directions ($\thetaegamma$, $\phiegamma$)
~\cite{angle def}
and emission time ($\tegamma$).
The offline event selection requires at least one $e^+$-track reconstructed in the spectrometer 
and pointing to the target, with minimal quality cuts applied.   
The blind region is defined by $48<\egamma<58$~MeV and $\left| \tegamma
\right|< 1~{\rm ns}$.

The positron track reconstruction in the spectrometer is based on a
Kalman filter technique~\cite{Kalman}. 
Effects of multiple scattering and energy loss in the detector materials 
in the presence of the non-uniform magnetic field are taken into account.   
Internal alignment of the DC is obtained by tracking cosmic ray muons without a magnetic field
and by minimizing the measured residuals in a manner independent of 
the initially assumed alignment~\cite{millipede}. 
The absolute position of the DC system is based on an optical survey. 

The magnetic field of the spectrometer was measured at the beginning of the experiment 
and only its major component along the beam axis is used in the analysis 
to avoid possible misalignment errors from the Hall probes; 
the other minor components are deduced from the derivatives of the measured primary component 
using Maxwell equations together with boundary conditions at a symmetry plane at the magnet centre where the minor components are nearly zero.
This magnetic field map agrees to within 0.2\% with the field computed for the geometry and currents of the spectrometer coils. 

The resolutions of the positron track direction are estimated
by exploiting tracks with two full turns in the DC. 
Each turn is treated as an independent track and the resolutions are
extracted from the difference between the two reconstructed sections. 
The energy resolution is evaluated 
by fitting the kinematic edge of the Michel decays 
and is well described by a sum of three Gaussians
with resolutions of 0.31\,MeV, 1.1\,MeV and 2.0\,MeV for the core (80\%)                                             
and the two tail (13\% and 7\%) components, respectively for 2009                                                 
and 0.32\,MeV, 1.0\,MeV and 2.0\,MeV for the core (79\%) and the two tail (14\% and 7\%)                         
components, respectively for 2010.                                                                                    

The decay vertex coordinates and the positron direction at the vertex are determined by extrapolating the reconstructed track back to the target.
The $\gamma$-ray direction is defined by the line connecting the decay vertex to 
the $\gamma$-ray conversion point measured by the LXe detector.

A geometrical correlation exists between
errors on 
$\phi_e$ 
at the vertex position and  $E_e$, which is measured by using the two-turn method
and is perfectly reproduced by the Monte Carlo (MC) simulation.
The $\phi_e$-resolution has a $\phi_e$-dependence due to the correlation and
has a minimum at $\phi_e = 0$,  where it is measured to be
$\sigma_{\phi_{\mathrm e}}=6.7 (7.2)$\,mrad for 2009 (2010) data 
\footnote{From here on we will quote in parentheses the value in the
2010 data when different from that in 2009.}.
The $\theta_{\mathrm e}$-resolution is measured  by the two-turn method to be
$\sigma_{\theta_{\mathrm e}} = 9.4 (11.0)$\,mrad.
The resolution on the decay vertex coordinates is also determined by the two-turn method;
along the beam axis it is described by a Gaussian with $\sigma_{z}=1.5 (2.0)$\,mm 
while in the vertical direction it is described by the sum of two Gaussians 
with $\sigma_{y}=1.1$\,mm for the core (87\%(85\%)) 
and $\sigma_{y}=5.3 (4.8)$\,mm for the tail.

The determination of the photon energy $\egamma$ in the LXe detector
is based on the sum of the number of scintillation photons detected by the
PMTs; correction factors take into account the different PMT geometrical acceptances.
Due to its geometry the detector response is not totally uniform over the photon entrance window; 
this is corrected for by using $\gamma$-lines from CW and CEX reactions.
The absolute energy scale and resolution at the signal energy $\egamma=52.8$\,MeV 
are determined by the CEX measurement; the resolution $\sigma_{\mathrm R}$, extracted from a Gaussian fit
to the high energy side of the spectrum, depends also on the depth ($w$) 
of the $\gamma-$ray conversion point from the photon entrance surface of the LXe detector: 
$\sigma_{\mathrm R}=1.9\% (w>2\,{\rm cm})$ and $2.4\% (w<2\,{\rm cm})$.  
The 3D-map of the measured resolutions is 
incorporated into the PDFs for the likelihood analysis.

The photon energy scale and the resolutions are cross-checked by fitting the 
background spectra measured in the side-bands with the theoretical RMD spectrum 
folded with the detector resolutions; 
the resolutions during the run are well represented by the CEX evaluations and 
the systematic uncertainty of the $\egamma$-scale is
estimated to be $\simeq 0.3\%$.
Since MEG operates at a high beam intensity, 
 it is important to recognize and unfold pile-up photons.
 For each event the spatial and temporal distributions of the PMT charge are studied 
 to identify photon pile-up in the LXe detector; 
 in case of positive identification, corrections to the PMT charges are applied.
 Cosmic ray events are rejected using their characteristic PMT charge distribution.

The position of the first interaction of the $\gamma$-ray in the LXe detector 
is derived from the light distribution measured by the PMTs close to the region of the energy deposition
by fitting the distribution with the expectation. 
The position resolution in the plane of the photon entrance window is measured to be $5\,$mm 
in a dedicated CEX run with a lead
slit-collimator placed in front of the LXe detector,
while the resolution along the depth $w$ of $6\,$mm and 
the position dependence of the resolutions are evaluated by a Monte Carlo simulation.

The resolutions on the relative directions ($\thetaegamma$, $\phiegamma$) 
are derived by combining the relevant resolutions of positrons and photons 
discussed above;
the results are 
14.5 (17.1)\,mrad for $\thetaegamma$ and 13.1 (14.0)\,mrad for $\phiegamma$.
The relative time $\tegamma$ is derived from the two time measurements
by the LXe detector and the TC, after correcting for the length of the particle flight-path.
The associated resolutions at the signal energy 
$146  (122)$\,ps
are evaluated from 
the RMD peak observed in  the $\egamma$ side-band; a
small correction takes into account the $\egamma$-dependence 
measured in the CEX calibration runs.
The position of the RMD-peak
corresponding to $\tegamma = 0$ was monitored constantly during the
physics data-taking period and found to be stable to within
$15$~ps.

A likelihood analysis is carried out for events in a portion of the blind region
(analysis region) 
defined by $48<\egamma<58\,$MeV, $50<\epositron<56\,$MeV,
$\left|\tegamma\right|<0.7\,$ns,
$\left|\thetaegamma\right|<50\,$mrad and
$\left|\phiegamma\right|<50\,$mrad. 
These intervals in the analysis variables are between five and twenty
sigmas wide to fully contain the signal events and also retain some background
events. 
The best estimates of the numbers of signal, RMD and accidental
background (BG) events in the analysis region are obtained by maximizing
the following likelihood function:

\begin{eqnarray}
\lefteqn{{\cal L}\left(\nsig, \nrd, \nbg\right) = } \nonumber \\
&&\frac {e^{- N}}{\nobs!} e^{ -\frac{(\nrd-\langle\nrd\rangle)^2}{2\sigma_{\rm RMD}^2} }
e^{ -\frac{(\nbg-\langle\nbg\rangle)^2}{2\sigma_{\rm BG}^2} }\times \nonumber  \\
&& \prod_{i=1}^{\nobs} \left( {\nsig} S(\vec{x}_i) + {\nrd}
R(\vec{x}_i) + {\nbg} B(\vec{x}_i) \right), \nonumber
\end{eqnarray}

where $\vec{x_i} = \left\{\egamma, \epositron, \tegamma,
\thetaegamma, \phiegamma \right\}$ is the vector of observables for
the $i$-th event, $\nsig$, $\nrd$ and $\nbg$ are the expected numbers of
signal, RMD and BG events, while $S$, $R$ and $B$ are
their corresponding PDFs. $N = \nsig + \nrd + \nbg$
and $\nobs(=311(645))$
is the observed total number of events in the
analysis window. $\langle\nrd\rangle(=27.2(52.2))$ and $\langle\nbg\rangle(=270.9(610.8))$ are the numbers of
RMD and BG events extrapolated from the side-bands together
with their uncertainties $\sigma_{\rm RMD}(=2.8(6.0))$ and $\sigma_{\rm BG}(=8.3(12.6))$, respectively.

The signal PDF $S(\vec{x_i})$ is the product of the PDFs for
$\epositron$, $\thetaegamma$, $\phiegamma$ and $\tegamma$, which are correlated variables, 
as explained above, 
and the $ \egamma$ PDF. 
The PDFs properly incorporate the measured resolutions and correlations 
among $\epositron, \thetaegamma$, $\phiegamma$ and $\tegamma$ 
on an event-by-event basis. 
The RMD PDF $R(\vec{x_i})$ is the product of the same
$\tegamma$-PDF as that of the signal and the PDF of the other four
correlated observables, which is formed by folding the theoretical
spectrum with the detector response functions.
The BG PDF
$B(\vec{x_i})$ is the product of the five PDFs, each of which is
defined by the single background spectrum, precisely measured in the
side-bands.
The dependence of the resolutions on the position of the
$\gamma$-ray interaction point and on the positron tracking quality
is taken into account in the PDFs.

A frequentist approach with a profile likelihood-ratio ordering~\cite{PDG, Feldman-Counsins} is used to compute the confidence intervals on $\nsig$:
\begin{eqnarray}
   \lambda_{\mathrm{p}} (\nsig) = \frac{{\cal L}(\nsig, \skew{3}\widehat{\widehat{N}}_{\rm RMD}(\nsig), \skew{3}\widehat{\widehat{N}}_{\rm BG}(\nsig))}{{\cal L}(\widehat{N}_{\rm sig}, \widehat{N}_{\rm RMD}, \widehat{N}_{\rm BG})},\nonumber
\end{eqnarray}
where the hat and double hat denote the best estimates maximizing the likelihood for floating and fixed $\nsig$, respectively.
Other, independent analysis schemes based on averaged PDFs without event-by-event information or Bayesian approach were also used 
and found to be compatible with
the analysis presented here to within 10 to 20\% difference in the obtained branching ratio upper limits.

In order to convert $\nsig$ into a branching ratio value
the normalization relative to the Michel decay is computed~\cite{MEG-NPB} 
by counting the number of Michel positrons passing
the same analysis cuts. This is accomplished by
means  of a pre-scaled Michel positron trigger enabled during the physics data-taking.
A correction to the pre-scaling factor due to positron pile-up in the TC is taken into account.
Another method for computing the normalization uses RMD events in the $\egamma$ side-band
and the theoretical branching ratio of the RMD. 
The normalizations calculated by these two independent methods are in good agreement 
and are combined to give the normalization factor with a 7\% uncertainty.

The sensitivity of the experiment with a null signal hypothesis is evaluated by
taking the median of the distribution of
the upper limit on the branching ratio obtained over an ensemble
of toy MC experiments. The rates of RMD and BG events, as measured
in the side-bands, are assumed in the simulated experiments.
The branching ratio sensitivity at 90\% confidence level (C.L.) is found to be
$3.3\times10^{-12}$ ($2.2\times10^{-12}$) for the 2009 (2010) 
data sample and $1.6\times10^{-12}$ when
2009 and 2010 are combined. These
sensitivities  are consistent with the upper limits
obtained by the likelihood analyses in several
comparable analysis regions of the $\tegamma$ side-bands.

\begin{figure}
\centering
\subfigure{%
\includegraphics[clip, width=.49\columnwidth]{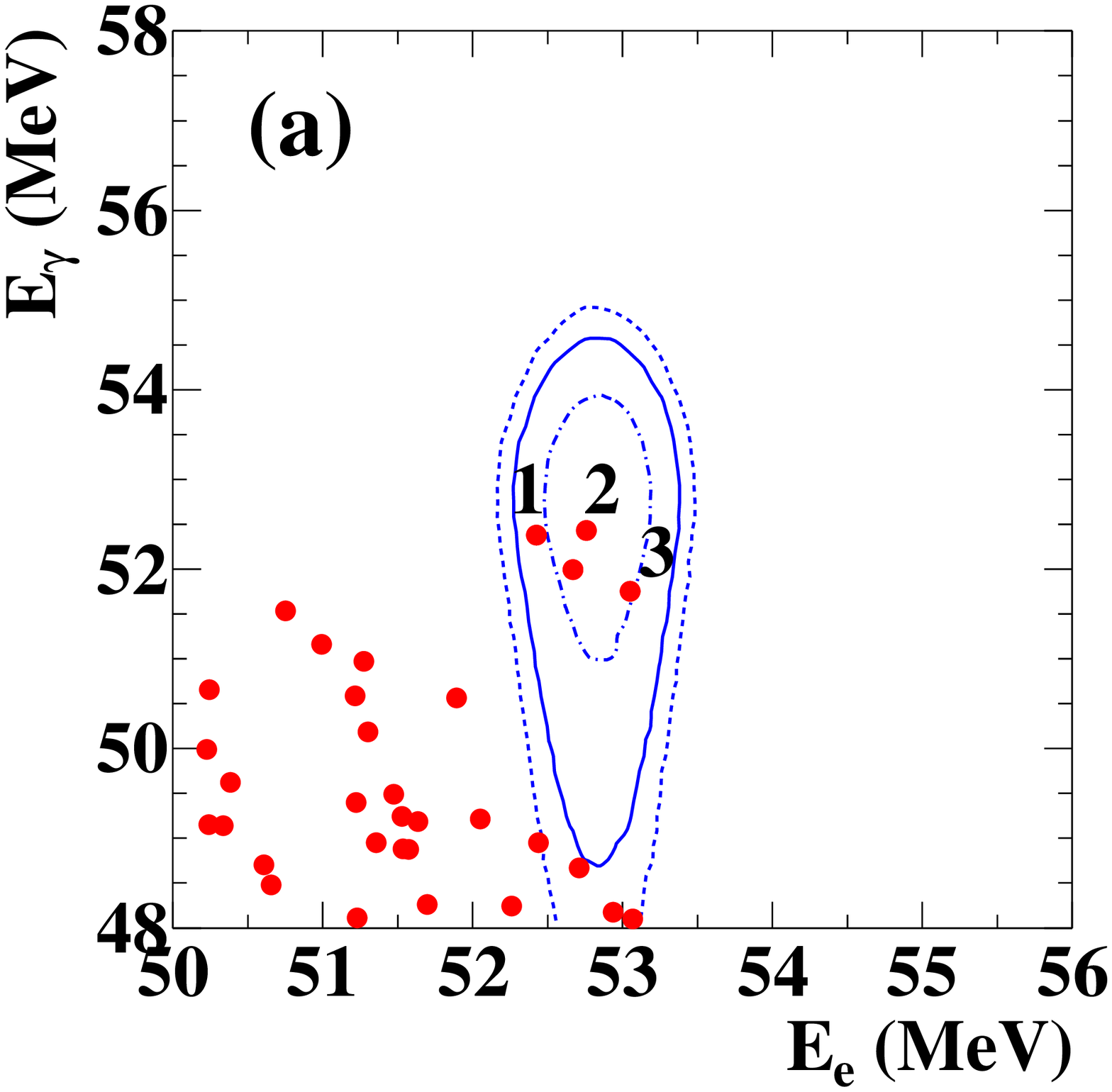}
}%
\subfigure{%
\includegraphics[clip, width=.49\columnwidth]{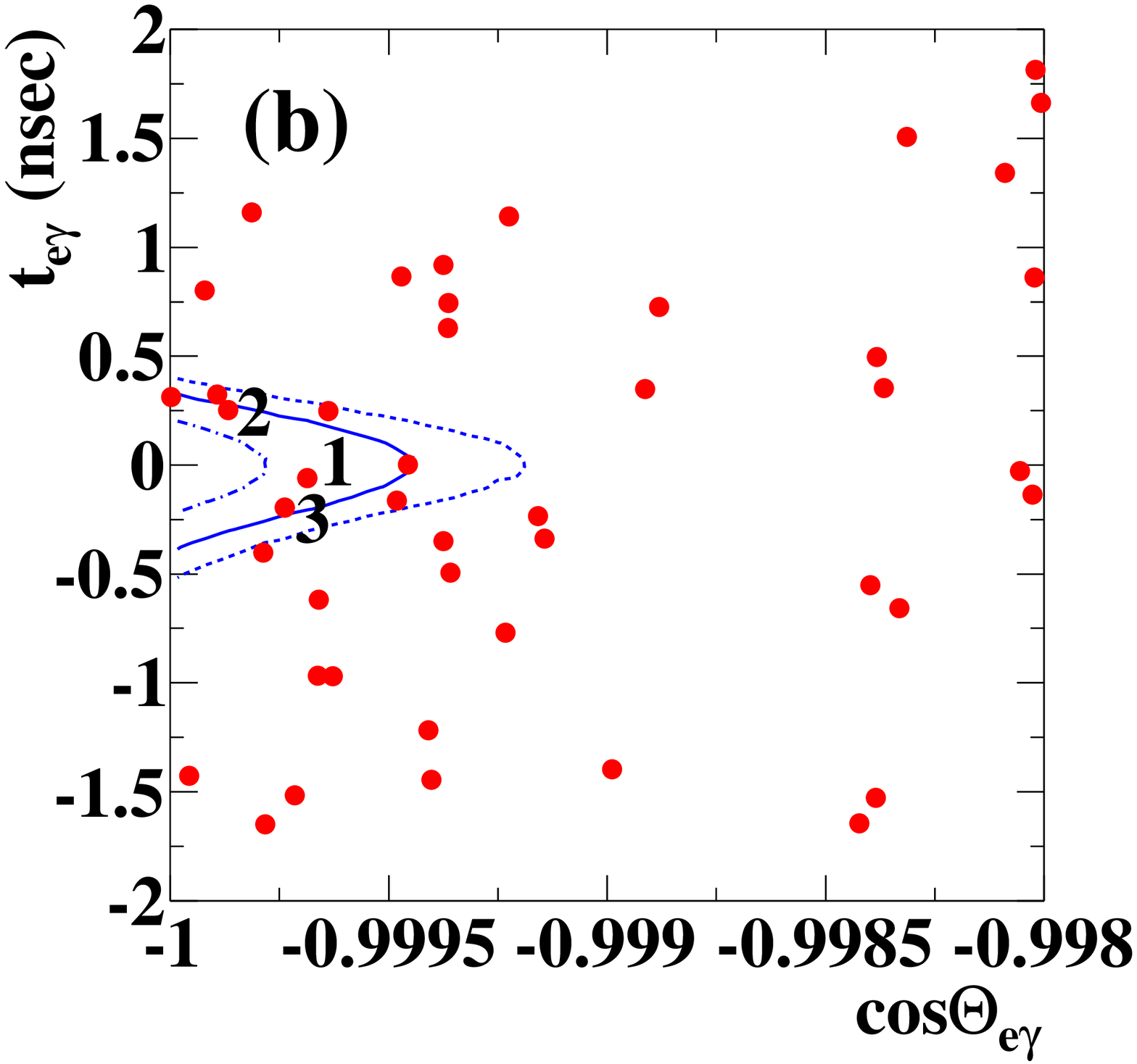}%
}
\subfigure{%
\includegraphics[clip, width=.49\columnwidth]{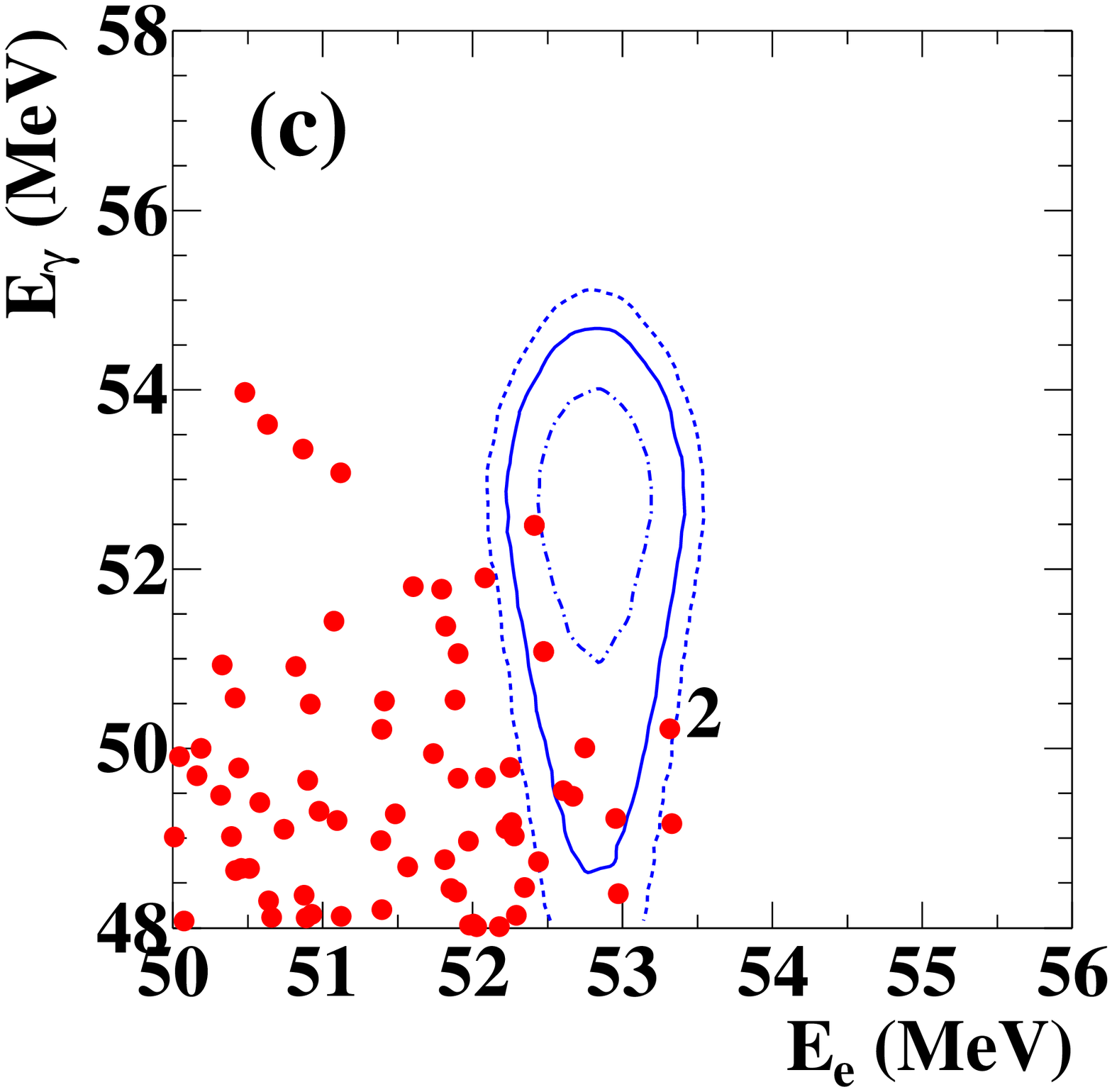}
}%
\subfigure{%
\includegraphics[clip, width=.49\columnwidth]{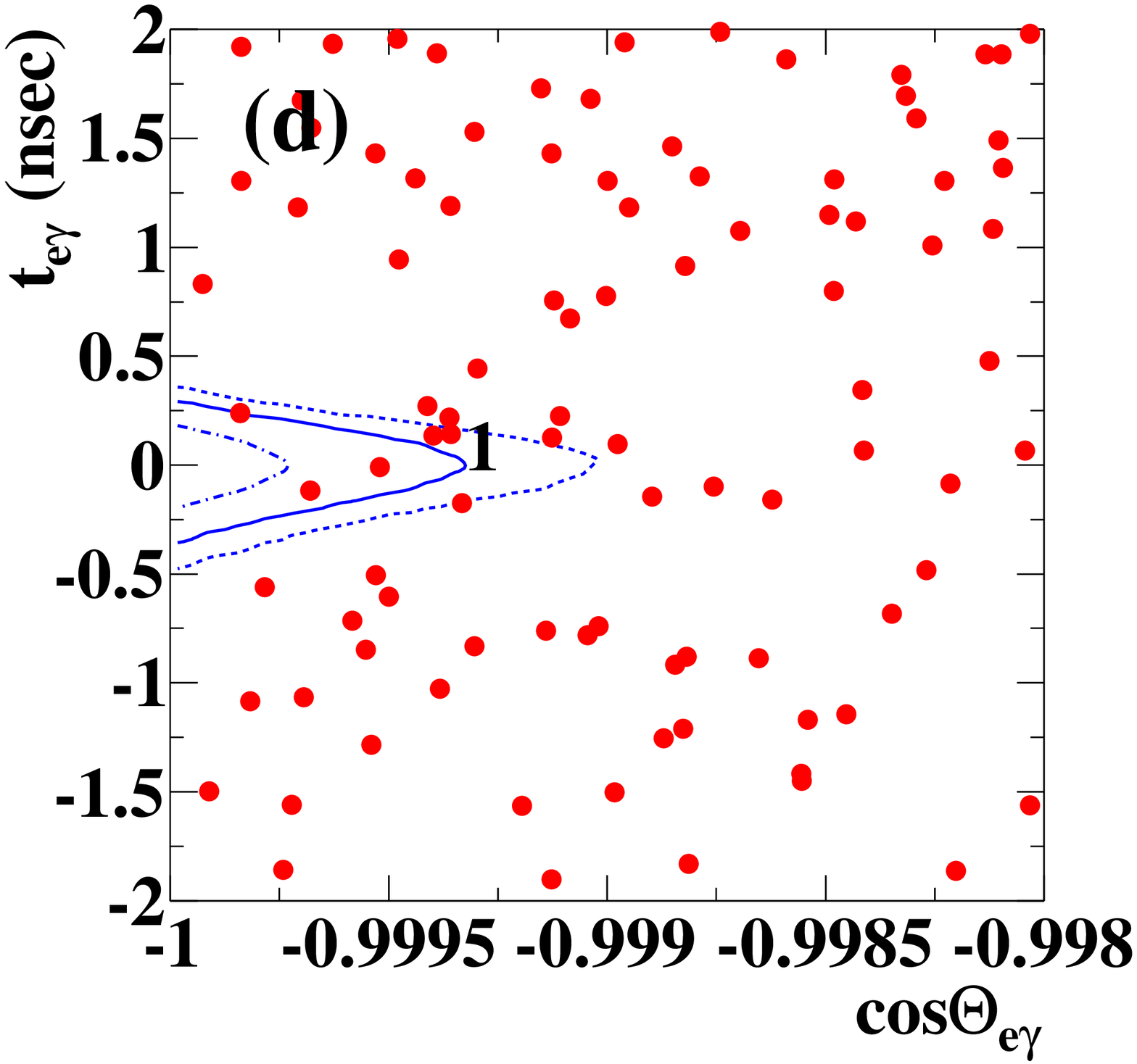}%
}
\caption{\label{fig:event distribution 2009/2010}
Event distributions in the analysis region
of (a) $\egamma$ vs $\epositron$ and (b) $\tegamma$  vs $\cos\Thetaegamma$ for 2009 data
and of (c) $\egamma$ vs $\epositron$ and (d) $\tegamma$  vs $\cos\Thetaegamma$ for 2010 data.
The contours of the PDFs (1-, 1.64- and 2-$\sigma$)
are shown, and a few events with the highest signal likelihood are numbered for each year. 
(The two highest signal likelihood events in 2010 data appear only in (c) or (d).) 
}
\end{figure}

After calibrations, optimization of the analysis algorithms
and background studies in the side-bands are completed, 
the likelihood analysis in the analysis region is performed.  In
Figures~\ref{fig:event distribution 2009/2010}
we present the distributions, for the 2009 and 2010 data samples respectively,
showing the events seen in the analysis region projected in the
$\egamma$ vs $\epositron$  and  $\tegamma$ vs $\cos\Thetaegamma$
planes, $\Thetaegamma$ being the opening angle between the
$\gamma$-ray and the positron.  
In plots $(a)$ and $(c)$ selections in $\tegamma$ and $\cos\Thetaegamma$, 
each of which is 90\% efficient on the signal, are 
applied ($|\tegamma|<0.28\,{\rm ns}$ and $\cos\Thetaegamma<-0.9996$)
; in plots $(b)$ and $(d)$
a selection in $\epositron$
which is 90\% efficient on the signal and a selection in $\egamma$
which is 73\% efficient on the signal inside the analysis window are applied
($52.3<\epositron<55\,{\rm MeV }$ and $51<\egamma<55\,{\rm MeV}$).
The contours of the signal PDF are
also drawn and a few events with the highest signal likelihood 
are numbered
in a decreasing order of relative signal likelihood,  
$S/(f_{\rm R}R+f_{\rm B}B)$,
$f_{\rm R} = 0.1$ and $f_{\rm B} = 0.9$ being
the fractions of the RMD and the BG measured in the side-bands, respectively.
High signal likelihood events were thoroughly checked and found to be randomly distributed in time and detector acceptance.

\begin{figure}[tb]
\begin{center}
\includegraphics[width=\columnwidth, clip=true]{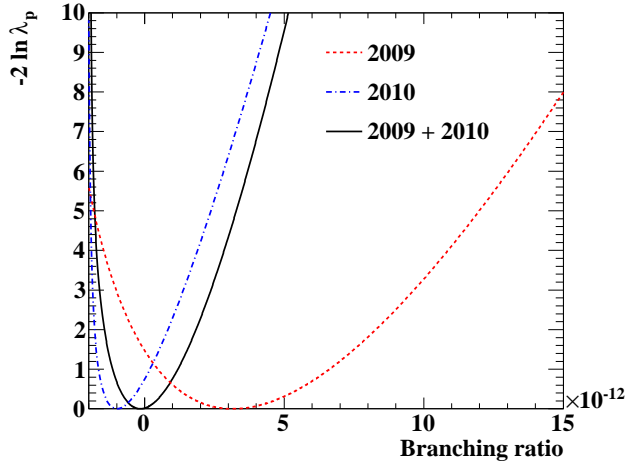}
\caption{\label{fig:likprofile}
Profile likelihood ratios as a function of the $\megsign$
branching ratio for 2009, 2010 and the combined 2009 + 2010 data sample. 
}
\end{center}
\end{figure}

The observed profile likelihood ratios as a function of the branching ratio for
2009, 2010 and the combined data sample are shown in Fig.~\ref{fig:likprofile}
~\cite{likelihood curve}.
The analysis of the full data sample gives a 90\% C.L. upper limit of $2.4\times 10^{-12}$,
which constitutes
the most stringent limit on the existence of the  $\megsign$ decay, 
superseding the previous limit by a factor of 5. 
The 90\% C.L. intervals as well as the best estimate of the branching ratio 
for 2009 and 2010 data separately 
are also given in Table~\ref{tab:BRtable}.
The 2009 data set, which gives a positive best estimate for the branching ratio, is consistent with the hypothesis $\BR = 0$ with an 8\% probability. 

The systematic uncertainties for the parameters of the PDFs and the
normalization factor are taken into account in the calculation of
the confidence intervals by fluctuating the PDFs 
according to the uncertainties. 
The largest contributions to the systematic uncertainty,
which amount to a shift of about 2\% in total in the branching ratio upper limit,
come from the uncertainties of
the offsets of the relative angles, the correlations in the positron observables and the normalization.

\begin{table}[tb]
\caption{\label{tab:BRtable} Best fit ($\BR_{fit}$), lower (LL) and upper limits (UL) 
at the 90\% C.L. of the branching ratio for the 2009, 2010 and combined 2009 + 2010 data sets.}
\begin{center}
\begin{tabular}{lccc}
\\{\bf Data set} & $\BR_{\rm fit}$ & LL & UL  \\[1mm] 
\hline
\hline\\[-3mm]

2009 & $\hspace*{\fill} 3.2\times 10^{-12}$\,\, &  $1.7 \times 10^{-13}$\,\, & $9.6\times 10^{-12}$  \\
2010 & $\hspace*{\fill} -9.9\times 10^{-13}$\,\, &  $-$\,\, & $1.7\times 10^{-12}$  \\
2009 + 2010 \,\, & $\hspace*{\fill} -1.5\times 10^{-13}$\,\, &  $-$\,\, & $2.4\times 10^{-12}$  \\

\hline \hline
\end{tabular}
\end{center}
\end{table}

The MEG experiment continues data-taking and is expected to 
explore the $\megsign$ decay down to a branching ratio sensitivity of a few times $10^{-13}$ in the next few years.

\section{Acknowledgements}
We are grateful for the support and co-operation provided by PSI as
the host laboratory and to the technical and engineering staff of
our institutes. This work is supported by DOE DEFG02-91ER40679
(USA), INFN (Italy) and MEXT KAKENHI 16081205 (Japan).


\begin{thebibliography}{99}
\bibitem{likelihood curve} These curves are not directly used to derive the upper limits 
which are obtained in a full frequentist approach as described above.
\end{thebibliography}
\end{document}